\documentclass[11pt]{article}

\newcounter{mnotecount}[section]

\renewcommand{\themnotecount}{\thesection.\arabic{mnotecount}}

\newcommand{\mnote}[1]
{\protect{\stepcounter{mnotecount}}$^{\mbox{\footnotesize
$
\bullet$\themnotecount}}$ \marginpar{
\raggedright\tiny\em
$\!\!\!\!\!\!\,\bullet$\themnotecount: #1} }

\usepackage{a4}
\usepackage{amssymb,latexsym}
\usepackage[mathscr]{eucal}
\usepackage{citesort}
\usepackage{url}
\usepackage{ntheorem}

\newtheorem{Theorem} {\sc  Theorem\rm} [section]
\newcommand{\tg}{\tilde{h}}%
\newcommand{\tx}{\tilde{x}}%
\newcommand{\tQ}{\widetilde{Q}}%

\newcommand{\tD}{\widetilde{D}}

\newcommand{\tR}{\widetilde{R}}
\newcommand{\tDel}{\widetilde{\Delta}}

\newcommand{\bea}{\begin{eqnarray}}

\newcommand{\eea}{\end{eqnarray}}
\newcommand{\bean}{\begin{eqnarray*}}

\newcommand{\eean}{\end{eqnarray*}}
\newcommand{\be}{\begin{equation}}

\newcommand{\ee}{\end{equation}}

\begin{document}
\title{Analyticity of strictly static and strictly stationary, inheriting and non-inheriting Einstein-Maxwell solutions}
\author{Paul Tod\thanks{{ E--mail}: \protect\url{paul.tod@st-johns.oxford.ac.uk}}
\\
Mathematical Institute and St John's College\\ Oxford}

\maketitle

\begin{abstract}
Following the technique of \cite{H1}, we show that strictly static and strictly stationary solutions of the Einstein-Maxwell equations are analytic in harmonic coordinates. This holds whether or not the Maxwell field inherits the symmetry.
\end{abstract}

\section{Introduction}
\label{intro}
In a classic paper from thirty seven years ago, \cite{H1}, M\"uller zum Hagen showed that static solutions of the Einstein vacuum equations are analytic in suitable (in fact harmonic) coordinates, at least where the staticity Killing vector is time-like (which we shall call \emph{strictly static}). He commented, without going into details, that the same should be true for the Einstein equations with sources, given restrictions on the kinds of source. In a second article, \cite{H2}, he showed that stationary solutions of the Einstein vacuum equations are also analytic in suitable coordinates, where the Killing vector is time-like (say \emph{strictly stationary}; in this article we shall only consider strictly static or strictly stationary solutions). It has been part of the folklore of General Relativity, though doesn't seem to have been published explicitly, that the Einstein-Maxwell equations with a Maxwell field inheriting the symmetry have this property. (Recall that the Maxwell field \emph{inherits} the symmetry if ${\mathcal{L}}_KF=0$ where $K$ is the Killing vector generating the symmetry and $F$ is the electromagnetic field tensor). In this note, we shall show that this is true, and that it is also true if the Maxwell field does not inherit the symmetry (i.e. when ${\mathcal{L}}_KF\neq 0$).

One would like to know whether the analyticity extends up to or even across any horizons, as this would assist black-hole classification theorems. Known static or stationary black-hole solutions all are analytic across the horizon but, aside from \cite{chr} for static vacuum solutions with non-degenerate horizons, one does not have any direct proofs that this must be so.

The basic technique in this article is to follow \cite{H1} and use the following theorem of Morrey \cite{mor}:

\begin{Theorem}[Morrey]
\label{thm2}
Given a system of second-order equations
$$\Phi^A(x^a,u^B,u^B_{\;\;,a},u^B_{\;\;,ab})=0$$
in an unknown $u^A$ for $A=1,2,\ldots ,N$, where $\Phi^A$ is analytic in all arguments and the system is elliptic in a domain $D$, then any $C^{2+\mu}$ solution in $D$ is in fact analytic.
\end{Theorem}
Recall that the system is elliptic if the symbol is non-degenerate, that is to say, define
\[L_B^{\;\;A}(x^a,p^a)=\sum_{a,b} p^ap^b\frac{\partial}{\partial u^B_{ab}}\Phi^A(x^a,u^B,u^B_{\;\;a},u^B_{\;\;ab})\]
and then we require
\[{\mathrm{det}}\; L_B^{\;\;A}(x^a,p^a)\neq 0,\]
for all $x^a$ in $D$ and all nonzero $p^a$.

The method of proof will be just as in \cite{H1}, that is we shall cast the Einstein equations into a form in which Theorem~\ref{thm2} can be applied. For strictly static, inheriting Einstein-Maxwell this is straightforward, but for the non-inheriting case there is a small complication. One of the Maxwell equations is first-order and so does not fit into the pattern considered by Theorem~\ref{thm2}. Consequently, we raise the order and obtain a system to which Theorem~\ref{thm2} does apply, and then recover the first-order equation via the Bianchi identities. In the strictly stationary case, we first assume that $a$ in equation (\ref{kv0}) below is constant - an example shows that this extra assumption is necessary - then we raise the order and obtain a system to which Theorem~\ref{thm2} does apply. This time we don't seem able to recover the first-order Maxwell equation but still we can argue as follows: any solution of the system including the first-order Maxwell equation satisfies the system with the order raised and therefore is analytic.

The plan of the article is as follows. Material on the static case is in section 2 and on the stationary case is in section 3. In sub-section 2.1, we review the static Einstein equations with a general stress-tensor. In sub-section 2.2, we review some fairly standard theory of the static inheriting Einstein-Maxwell equations as it may be found for example in \cite{skmhh}, and note that the equations are already in a form to which Theorem~\ref{thm2} can be applied. Thus these solutions are analytic. Then in sub-section 2.3, and drawing on \cite{me1}, we recall the theory of non-inheriting Einstein-Maxwell equations and see how the argument above, first raising the order and then recovering the first-order equation, goes through. The corresponding programme is carried through for stationary solutions in section 3.

\section{The strictly static case}
\subsection{The Einstein equations}
In \cite{H1}, M\"uller zum Hagen first shows that comoving coordinates can be introduced:

\begin{Theorem}[M\"uller zum Hagen]
\label{lem1}

If a static space-time $M$ is $C^5$, with the Killing vector $K$ being $C^4$, and the metric $C^3$, then there is a $C^4$ atlas with
\[
g=F(x^k)dt^2-h_{ij}(x^k)dx^idx^j
\]
with $F$ and $h_{ij}$ both $C^3$.
\end{Theorem}
We shall write $a,b,\ldots$ for space-time indices and $i,j,k,\ldots$ for spatial indices.

In the standard way (see e.g. \cite{skmhh}), for a convenient form of Einstein's equations, we set $F=e^{2U}$, which is why we insist on strict staticity, and introduce the rescaled metric:
\[
\tg_{ij}=e^{2U}h_{ij}.
\]
The (three-dimensional) Ricci tensor transforms as 
\[\tR_{ij}=R_{ij}-D_iD_jU+U_{,i}U_{,j}-h_{ij}(\Delta U+|DU|^2),\]
where $D_i$ is the Levi-Civita derivative for the metric $h_{ij}$ and $\Delta=h^{ij}D_iD_j$ is its Laplacian. The Einstein field equations can be written in terms of $h_{ij}$ as the system
\bean
R_{ij}&=&D_iD_jU+U_{,i}U_{,j}+P_i^aP_j^b(T_{ab}-\frac{1}{2}Tg_{ab})\\
\Delta U+|DU|^2&=&e^{-2U}K^aK^b(T_{ab}-\frac{1}{2}Tg_{ab}),
\eean
where $P_i^a=g_i^a$ is the projection into surfaces of constant $t$ and $K^a$ is the Killing vector. Then, in terms of the rescaled metric $\tg_{ij}$ the Einstein equations become
\bea
\tR_{ij}&=&2U_{,i}U_{,j}+(P_i^aP_j^b-\tg_{ij}e^{-4U}K^aK^b)(T_{ab}-\frac{1}{2}Tg_{ab})\label{fe3}\\
\tDel U&=&e^{-4U}K^aK^b(T_{ab}-\frac{1}{2}Tg_{ab}),\label{fe4}
\eea
where $\tDel$ is the Laplacian for $\tg_{ij}$. 

Still following \cite{H1}, we introduce harmonic coordinates $\tx^{i'}$, $i=1,2,3$ for $\tg_{ij}$ satisfying:
\[\tg^{-1/2}(\tg^{1/2}\tg_{mn}x^{i'}_{,m})_{,n}=0\]
where $\tg=\mathrm{det}(\tg_{ij})$, then (dropping primes)
\be
\tR_{ij}=\frac{1}{2}\tg^{mn}\tg_{ij,mn}+\tQ_{ij},
\label{fe5}
\ee
where $\tQ_{ij}$ does not contain second derivatives of $\tg_{ij}$. 

The vacuum Einstein equations can be obtained from (\ref{fe3})-(\ref{fe4}) by dropping all terms in $T_{ab}$. In harmonic coordinates, with the aid of (\ref{fe5}), it is easy to check that they are in the form suitable for application of Theorem~\ref{thm2} and so, following \cite{H1}, we conclude that strictly static solutions of the vacuum equations are analytic in these coordinates, which by their definition are only available away from any horizons (where $F=0$). We seek next to carry through the corresponding calculation for the Einstein-Maxwell equations.

\subsection{Inheriting Einstein-Maxwell}
In this section, we review some standard and fairly familiar material on the inheriting Einstein-Maxwell equations. This is in order to extend the result of the previous section to this case, but also to prepare the way for the non-inheriting case in the next section.

Given a Maxwell field $F_{ab}$ which inherits the symmetry, we introduce the electric and magnetic fields in the usual way as
\be
E_a=F_{ab}K^b\;,\;\;B_a=\star F_{ab}K^b.
\label{e1}
\ee
(though some authors including \cite{me1} may put a factor $e^{-U}$ on the right in these definitions). Inheritance of the symmetry is expressed as
\[{\mathcal{L}}_KE_a=0,\;\;{\mathcal{L}}_KB_a=0,\]
and, as is very familiar, implies that $E_a$ and $B_a$ are both closed. However, for a static Killing vector, one may without loss of generality set $B_a=0$ (since staticity implies that the Poynting vector vanishes, so that $E_a$ and $B_a$ are proportional; the Maxwell equations then force the function of proportionality to be a constant and then a duality rotation sets $B_a$ to zero). 

In a simply-connected region, closedness of $E_a$ implies that there exists a potential $\phi$ with $E_a=\phi_{,a}$. The Maxwell tensor is 
\[F_{ab}=2e^{-2U}E_{[a}K_{b]},\]
and, since $K^a$ is a static Killing vector we also have
\[\nabla_aK_b=2U_{[,a}K_{b]}.\]
The remaining Maxwell equation is now
\[\nabla_aE^a=2E^aU_{,a},\]
or 
\[\nabla_a(e^{-2U}E^a)=0.\]
In terms of $\phi$ this is
\be
\nabla^a(e^{-2U}\phi_{,a})=0.
\label{m1}
\ee
The energy-momentum tensor is
\bean
T_{ab}&=&2F_{ac}F^c_{\;\;b}+\frac{1}{2}F_{cd}F^{cd}g_{ab}\\
&=&-2e^{-2U}E_aE_b+2e^{-4U}|E|^2K_aK_b-g_{ab}e^{-2U}|E|^2,
\eean
where $|E|^2=-g^{ab}E_aE_b$ which is positive with our conventions.
Substitute into (\ref{fe3}) and (\ref{fe4}) to find
\bea
\tR_{ij}&=&2U_{,i}U_{,j}-2e^{-2U}\phi_{,i}\phi_{,j}\label{ie1}\\
\tDel U&=&e^{-2U}\tg^{ij}\phi_{,i}\phi_{,j}\label{ie2}
\eea
and the Maxwell equation (\ref{m1}) becomes
\be
\tDel\phi=2\tg^{ij}U_{,i}\phi_{,j}.\label{ie3}
\ee
The strictly static, inheriting Einstein-Maxwell equations have been reduced to the system (\ref{ie1})-(\ref{ie3}). These equations are quite familiar and may be found for example on p285 of \cite{skmhh}. In harmonic coordinates, the equations are second-order elliptic and then by Theorem~\ref{thm2} their solutions are analytic.

\subsection{Non-inheriting Einstein-Maxwell}
In this section, we shall follow the derivations in \cite{me1} but with a slightly different definition of $E_a$, $B_a$ and $W_a$ from that reference. If the static symmetry is not inherited, then, as shown in \cite{me1}, without loss of generality, the Maxwell tensor satisfies 
\be
{\mathcal{L}}_KF_{ab}=-a\star F_{ab},\label{kv0}
\ee
for real, non-zero constant $a$. With $E_a$ and $B_a$ defined as before, this gives
\[{\mathcal{L}}_K E_a=-aB_a\;;\;\;{\mathcal{L}}_KB_a=aE_a.\]
We may therefore introduce $W_a$ by
\[E_a=W_a\sin(at)\;;\;\;B_a=-W_a\cos(at),
\]
to find
\[
{\mathcal{L}}_KW_a=0.
\]
The Maxwell equations become
\bea
\nabla^b(e^{-2U}W_b)&=&0\label{w6}\\
\epsilon^{ab}_{\;\;\;\;cd}K^d\nabla_aW_b&=&-aW_c,\label{w7}
\eea
and, in terms of $W_a$, the energy-momentum tensor is
\be
T_{ab}=-2e^{-2U}W_aW_b+2e^{-4U}|W|^2K_aK_b-e^{-2U}|W|^2g_{ab}.
\label{t2}
\ee
From (\ref{fe3}) and (\ref{fe4}), with (\ref{w6}), (\ref{w7}) and (\ref{t2}), we obtain the system
\bea
\tR_{ij}&=&2U_iU_j-2e^{-2U}W_iW_j\label{r1}\\
\tDel U&=&e^{-2U}\tg^{ij}W_iW_j\label{v1}\\
\tilde{\epsilon}_i^{\;\;jk}\tD_jW_k&=&ae^{-2U}W_i.\label{w3}
\eea
When we rescale, we assume that $W_i$ does not change; thus we must be careful to distinguish $h^{ij}W_j$ from $\tg^{ij}W_j$. For the rest of this section, indices are raised and lowered only with $\tg$.

The problem with the system just obtained is that equation (\ref{w3}) for $W_i$ is first-order, so that Theorem~\ref{thm2} does not immediately apply. To deal with this, we decompose $W_i$ into closed and co-closed parts:
\be
W_i=\phi_{,i}+\psi_i,
\label{w4}
\ee
where $\tD_i\psi^i=0$. Then from the divergence of (\ref{w3}),
\be
\tDel \phi=2\tg^{ij}U_{,i}(\phi_{,j}+\psi_j),
\label{w5}
\ee
which is a suitable, second-order elliptic equation for $\phi$. The rest of (\ref{w3}) can be written
\be
e^{2U}\tilde{\epsilon}_i^{\;\;jk}\tD_j\psi_k=a(\phi_{,i}+\psi_i).
\label{psi1}
\ee
Apply $e^{2U}\tilde{\epsilon}_k^{\;\;mi}\tD_m$ to (\ref{psi1}) and simplify to obtain
\be
\tDel\psi_k=(\tD_k\tD_m-\tD_m\tD_k)\psi^m+2U_{,m}\tD_k\psi^m-2\tg^{mn}U_{,m}\tD_n\psi_k+a^2(\phi_{,k}+\psi_k),
\label{psi2}
\ee
where indices are raised and lowered with $\tg$. The first term on the right simplifies with the Ricci identity
\[(\tD_k\tD_m-\tD_m\tD_k)\psi^m=-\tR_{km}\psi^m,\]
 and now we have a system consisting of (\ref{r1}), (\ref{v1}), (\ref{w5}) and (\ref{psi2}) to which Theorem~\ref{thm2} can be applied. This system will have analytic solutions in harmonic coordinates, which is sufficient for our purposes: non-inheriting, strictly static Einstein-Maxwell solutions necessarily satisfy the system including (\ref{psi2}) and so are analytic. 

It is worth noting that we can in fact recover the first-order equation (\ref{w3}) for $W_i$. To accomplish this, introduce the notation:
\[L_i^{\;\;j}:=e^{2U}\tilde{\epsilon}_i^{\;\;mj}\tD_m,\]
so that (\ref{psi1}) and (\ref{psi2}) are respectively
\bea
L_i^{\;\;j}\psi_j&=&a(\phi_{,i}+\psi_j)\nonumber\\
L_k^{\;\;i}L_i^{\;\;j}\psi_j&=&a^2(\phi_{,i}+\psi_j).\label{psi4}
\eea
Now introduce $\chi_j$ by
\[\chi_j=L_j^{\;\;k}W_k-aW_j,\]
so that (\ref{w3}) is just $\chi_j=0$. From (\ref{psi4}) we have
\be
L_i^{\;\;j}\chi_j+a\chi_i=0
\label{chi2}
\ee
so that either $\chi_j=0$, which is (\ref{w3}), or $\chi_i$ is a non-zero solution of (\ref{chi2}), which is just (\ref{w3}) with the sign switched on $a$. Now return to the system (\ref{r1})-(\ref{v1}) and calculate the contracted Bianchi identity. This turns out to be
\be
\tD^i(\tR_{ij}-\frac{1}{2}\tR\tg_{ij})=2e^{-4U}\tilde{\epsilon}_j^{\;\;ik}W_i\chi_k,\label{cbi}\ee
where we have used 
\[\tD_iW^i-2U_{,i}W^i=0,\]
which is permissible since, given (\ref{w4}) this is just (\ref{w5}) which is part of the system already solved.

Given the (analytic) solution of the system, the left-hand-side of (\ref{cbi}) must vanish. Therefore so must the right-hand-side and either $\chi_j=0$, in which case (\ref{w3}) is satisfied, or $\chi_i$ is proportional to $W_i$, say $\chi_i=fW_i$ for some function $f$. In this second case we obtain
\bean
0&=&L_i^{\;\;j}\chi_j+a\chi_i\\
&=&L_i^{\;\;j}(fW_j)+afW_i\\
&=&f(\chi_i+aW_i)+afW_i+W_jL_i^{\;\;j}f\\
&=&f(f+2a)W_i+W_jL_i^{\;\;j}f,
\eean
using (\ref{chi2}) and the definition of $\chi_i$. Now contract with $W^i$ and use the definition of $L_i^{\;\;j}$ to find
\[f(f+2a)W^iW_i=0.\]
If $W_i$ isn't zero then $f=-2a$ (since by assumption for this part neither $\chi_i$ nor $f$ is zero) but now
\[\chi_i:=L_j^{\;\;k}W_k-aW_j=-2aW_i,\]
so that
\[L_j^{\;\;k}W_k+aW_j=0,\]
and we recover (\ref{w3}) but with the sign switched on $a$. Since $a$ enters the second-order system we are solving (specifically (\ref{psi2})) only through its square, we should expect to have this ambiguity in the first-order equation (\ref{w3}).

\section{The stationary case}
We now follow the corresponding chain of arguments to see that, in the strictly stationary case too, first vacuum, then inheriting Einstein-Maxwell, and then non-inheriting Einstein Maxwell solutions, subject to an extra assumption, are analytic in harmonic coordinates.

\subsection{The Einstein equations}

We begin with the result from \cite{H2} on comoving coordinates in stationary space-times corresponding to theorem~\ref{lem1}:
\begin{Theorem}[M\"uller zum Hagen]
\label{lem2}

If a stationary space-time $M$ is $C^5$, with the Killing vector $K$ being $C^4$, and the metric $C^3$, then there is a $C^4$ atlas with
\[ 
g=F(dt+A_idx^i)^2-h_{ij}dx^idx^j
\]
with $F$, $A_i$ and $h_{ij}$ all $C^3$.
\end{Theorem}
As before, to simplify the Einstein equations we put $F=e^{2U}$ and $\tg_{ij}=e^{2U}h_{ij}$, and introduce
\be
\omega_i=e^U\epsilon_i^{\;\;jk}\partial_jA_k.
\label{met3}
\ee
Our definition of $\omega$ differs slightly from the definition in \cite{skmhh}: if the definition there is $\omega_i^{ES}$ then $\omega_i=e^{-2U}\omega_i^{ES}$. Note that
\be
D_i\omega^i-\omega^iU_i=0.
\label{es3}
\ee
There is more choice in the coordinate $t$ in the stationary case than in the static case. To reduce it, and following \cite{H2}, we impose the condition that $t$  be a harmonic coordinate, and then 
\be
g^{ab}\nabla_a\nabla_bt=e^{-U}D_i(e^Uh^{ij}A_j)=0.
\label{met4}
\ee
The Einstein field equations become
\bean
R_{ij}&=&D_iD_jU+U_iU_j+\frac{1}{2}(\omega_i\omega_j-h_{ij}\omega_k\omega^k)\nonumber\\
&&+P_i^aP_j^b(T_{ab}-\frac{1}{2}Tg_{ab})\\
\Delta U+|DU|^2+\frac{1}{2}h^{ij}\omega_i\omega_j&=&e^{-2U}K^aK^b(T_{ab}-\frac{1}{2}Tg_{ab})\\
\epsilon_i^{\;\;jk}D_j(e^{2U}\omega_k)&=&2e^{3U}P_i^aK^b(T_{ab}-\frac{1}{2}Tg_{ab})
\eean
These can be checked against the form of the equations given in \cite{skmhh}, taking account of the different conventions there.

For stationary vacuum solutions, we set the stress-tensor terms to zero and write the equations in terms of the rescaled metric $\tg_{ij}$ to find 
\bea
\tR_{ij}&=&2U_{,i}U_{,j}+\frac{1}{2}\omega_i\omega_j\label{ref1}\\
\tDel U&=&-\frac12\tg^{ij}\omega_i\omega_j\label{ref2}\\
\tilde{\epsilon}_i^{\;\;jk}(\tD_j\omega_k+2U_{,j}\omega_k)&=&0\label{ref3}
\eea
(here $\omega_i$ and in the next set of equations $A_i$ are both assumed not to change under rescaling) while (\ref{met3}), (\ref{es3}) and (\ref{met4}) become
\bea
\omega_i&=&e^{2U}\tilde{\epsilon}_i^{\;\;jk}\partial_jA_k\label{ref4}\\
\tg^{ij}\tD_i(e^{-2U}\omega_j)&=&0\label{ref5}\\
\tD_i(\tg^{ij}A_j)&=&0.\label{ref6}
\eea
In harmonic coordinates, the system consisting of (\ref{ref1}) and (\ref{ref2}) is in a suitable form to apply Theorem~\ref{thm2} but (\ref{ref3}) isn't. We write it instead as an equation on $A_i$ using (\ref{ref4}) to find
\[
\tg^{ij}(\tD_i\tD_jA_k+4U_i(\tD_jA_k-\tD_kA_j)-(\tD_i\tD_k-\tD_k\tD_i)A_j)=0,
\]
when, with the aid of the Ricci identity, it takes the right form to apply Theorem~\ref{thm2}. Now, following \cite{H2}, we conclude that strictly stationary vacuum solutions are analytic in harmonic coordinates.

\subsection{Inheriting Einstein-Maxwell}
For inheriting Einstein-Maxwell, we introduce the electric and magnetic fields as in (\ref{e1}) and then both are closed, so that
\[E_a=F_{ab}K^b=\nabla_a\alpha,\;\;B_a=\star F_{ab}K^b=\nabla_a\beta,\]
for two scalar potentials $\alpha$ and $\beta$. In rescaled variables, the Einstein-Maxwell equations become
\bean
\tR_{ij}&=&2U_{,i}U_{,j}+\frac{1}{2}\omega_i\omega_j-2e^{-2U}(\alpha_{,i}\alpha_{,j}+\beta_{,i}\beta_{,j})\\
\tDel U&=&\frac12\tg^{ij}(2e^{-2U}(\alpha_{,i}\alpha_{,j}+\beta_{,i}\beta_{,j})-\omega_i\omega_j)\\
\tilde{\epsilon}_i^{\;\;jk}(\tD_j\omega_k+2U_{,j}\omega_k)&=&-4e^{-2U}\tilde{\epsilon}_i^{\;\;jk}\alpha_{,j}\beta_{,k}\\
\tDel\alpha&=&\tg^{ij}(2U_{,i}\alpha_{,j}+\omega_i\beta_{,j})\\
\tDel\beta&=&\tg^{ij}(2U_{,i}\beta_{,j}-\omega_i\alpha_{,j}),
\eean
together with (\ref{ref4})-(\ref{ref6}).

As for vacuum, we eliminate $\omega_i$ in favour of $A_i$, then in harmonic coordinates the system is elliptic, Theorem~\ref{thm2} applies and solutions are analytic.

\subsection{Noninheriting Einstein-Maxwell}
In the stationary but non-static, non-inheriting case, we still have (\ref{kv0}) but it is now an extra assumption that $a$ be constant. (For this, see the discussion in \cite{me1}: it is possible to have a non-constant $a$ with a null Maxwell field; there are pp-wave examples which are stationary, non-static and non-analytic in harmonic coordinates, for example the metric
\[g=2du(dv+A^2\zeta\overline\zeta du)-2d\zeta d\overline\zeta,\]
with Maxwell field $F=Ae^{if(u)}du\wedge d\overline\zeta + {\mathrm{c.c.}}$ for real constant $A$ and non-constant but non-analytic $f(u)$; the stationarity Killing vector is $K=\partial/\partial u$ and ${\mathcal{L}}_KF=f'\star F$ so that $a=-f'$). 

We shall therefore assume that $a$ is constant. We introduce $E_a$ and $B_a$ as before, though these are not now closed. We find
\[E_a+iB_a=\zeta_ae^{iat},\]
where $\zeta_a$, the counterpart of $W_a$ in section 2.3, is complex and ${\mathcal{L}}_K\zeta_a=0$. We may decompose $\zeta_a$ into real and imaginary parts as
\[\zeta_a=\xi_a+i\eta_a.\]
There is a residual `gauge freedom' under change of choice of time-coordinate:
\[t'=t+f(x^i),\;\zeta'_a=e^{-iaf}\zeta_a,\;A'_a=A_a-f_{,a}.\]
The Maxwell equations split into divergence equations which can be written in the form
\[D_i\zeta^i-(U_{,i}-i\omega_i+iaA_i)\zeta^i=0,\]
and curl equations:
\[\epsilon_i^{\;\;jk}(D_j\zeta_k-iaA_j\zeta_k)=ae^{-U}\zeta_i.\]
In the rescaled variables, these are:
\be
\tD_i(\tg^{ij}\zeta_j)-\tg^{ij}(2U_{,i}-i\omega_i+iaA_i)\zeta_j=0
\label{div1}
\ee
and
\be
\tilde\epsilon_i^{\;\;jk}(\tD_j\zeta_k-iaA_j\zeta_k)=ae^{-2U}\zeta_i.
\label{cur1}
\ee
where $\zeta_i$ is assumed not to change under rescaling. It is straightforward to check that the divergence equations follow from the curl equations, and that both sets retain the residual gauge invariance. The field equations are the Einstein equations in the rescaled variables:
\bea
\tR_{ij}&=&2U_{,i}U_{,j}+\frac{1}{2}\omega_i\omega_j-2e^{-2U}(\xi_i\xi_j+\eta_i\eta_j)\label{rne1}\\
\tDel U&=&\frac12\tg^{ij}(2e^{-2U}(\xi_i\xi_j+\eta_i\eta_j)-\omega_i\omega_j)\label{rne2}\\
\tilde{\epsilon}_i^{\;\;jk}(\tD_j\omega_k+2U_{,j}\omega_k)&=&-4e^{-2U}\xi_j\eta_k\label{rne3}
\eea
together with the Maxwell equations. We need to raise the order of the first-order Maxwell equations, (\ref{div1}) and (\ref{cur1}), as before, so decompose $\zeta_i$ into closed and co-closed pieces:
\[\zeta_i=\nu_{,i}+Z_i\]
for $\tD$-co-closed $Z_i$: $\tD_iZ^i=0$. 
By imposing the divergence equations (\ref{div1}), we obtain the following elliptic equation for $\nu$:
\be
\tDel \nu=\tg^{ij}(2U_{,i}+i(aA_i-\omega_i))(\nu_{,j}+Z_j).
\label{rne4}
\ee
%
We shall raise the order on the curl equations (\ref{cur1}). Set
\be
\Lambda_i=e^{2U}\tilde\epsilon_i^{\;\;jk}(\tD_j\zeta_k-iaA_j\zeta_k),
\label{rne6}
\ee
 then for the new second-order equation, we impose
\be
e^{2U}\tilde\epsilon_i^{\;\;jk}(\tD_j\Lambda_k-iaA_j\Lambda_k)=a^2\zeta_i.
\label{rne7}
\ee
which is a second-order equation for $Z_i$, in which the highest-order derivative terms are
\[e^{4U}(\tg^{jm}(\tD_j\tD_i-\tD_i\tD_j)Z_m-\tg^{jl}\tD_j\tD_lZ_m).\]
Here, as in (\ref{psi2}), the Ricci identity allows the first part to be expressed without derivatives, while the second part is in the right form for Theorem~\ref{thm2}.

Note that (\ref{rne7}) preserves the gauge-invariance and, if (\ref{cur1}), holds then $\Lambda_i=a\zeta_i$ and so (\ref{rne7}) holds. Thus any solution of the system (\ref{div1})-(\ref{rne3}) also satisfies the system (\ref{rne1})-(\ref{rne4}) together with (\ref{rne7}). This latter system, with $\omega_i$ eliminated in favour of $A_i$, is second-order elliptic and so Theorem~\ref{thm2} can be applied. We conclude that strictly stationary, non-inheriting solutions of the Einstein-Maxwell equations with constant $a$ are analytic in harmonic coordinates.

It doesn't seem so straightforward in the stationary case as in the static to recover the first-order equation (\ref{cur1}) (note that the ambiguity in the sign of $a$ which was a feature of recovering the first-order equation in the static case, is not present in the stationary case since, for example, (\ref{rne4})-(\ref{rne7}) all contain $a$  to the first power). This difficulty seems to be connected to the fact that there are six real equations in (\ref{cur1}) but one obtains only three real equations from the contracted Bianchi identity.

 \section*{Acknowledgement}
I am grateful to Piotr Chru\'sciel for useful conversations.

%
\end{document}